\documentclass[12pt]{article}
\usepackage{myown}
\newcommand{\etal}{{\it et al}}
\title{\large \sf \begin{flushright} Report 
     NPI \v{R}e\v{z}--TH--01/2000
    \end{flushright}
\vspace{5mm}
 \bf Tables of internal conversion coefficients for superheavy
    elements}
\author{\normalsize M.Ry\v sav\'y\thanks{e-mail: rysavy@ujf.cas.cz}
   \ and  O.Dragoun \\
\small \it Nuclear Physics Institute, Acad. Sci. of Czech Republic,
    \vspace{-0.7ex}\\
\small \it CZ-250 68 \v Re\v z near Prague, Czech Republic}
\date{ }

\begin{document}

\maketitle

\section{Introduction}
At present, the attempt to synthetize superheavy elements is
one of important tasks of modern physics. The study of the
structure of such isotopes is then a question of a near
future.

Very recently, in the isotope of $^{254}$No (Z=102), the ground-state
 band of even--even nucleus up to spin 14 was identified \cite{Rei99}.
Moreover a production of more heavier element -- Z=118, A=293 -- was
indicated \cite{Nin99}. Then a further development of $\gamma$--ray
spectroscopy can be expected. However, for such high Z's, 
the internal conversion
strongly prevails over the emission of $\gamma$--rays and must 
be taken into account. In this work we present the internal conversion
coefficients (ICC) for superheavy elements 104$\leq$Z$\leq$126.

\section{Calculations}
The ICC were calculated using the computer program NICC \cite{Ry77}.
The program solves the Dirac equation for both bound and free 
electron states using the formulae by B\"{u}ring \cite{Bue65}.
Then it performs direct integration with reasonably
small step to obtain the conversion matrix elements. The atom
is described by a Hartree--Fock--Slater potential, the nucleus
is expected to bear the Fermi charge distribution. In this work,
we use the potential of Lu \etal\ \cite{Lu71}.

Usually, the kinetic energy of the converted electron is derived
from experimental binding energy of that electron prior to
conversion. Due to the absence of the experimental data, we use
the eigenvalues from \cite{Lu71} instead. Those eigenvalues are
known to be close to the experimental binding energies where
available (within tens of eV for inermost shells, within eV's
for the outermost shells). Moreover, the dependence of the
theoretical ICC on the kinetic energy of the emitted electron
(except if this energy is close to zero, i.e.  for transition
energies near to the threshold for the particular subshell)
is not too strong.

For Z=104 only, we can compare our ICC with those of R\"{o}sel
 \etal\ \cite{Roe78} and Band and Trzhaskovskaya \cite{Ba78}.
For comparison, we have chosen the former ones since they
are calculated using almost the same atomic model as we used.
In most cases, the agreement is better than 1 percent, only
exceptionally within 2 -- 3 percent.
\section{Arrangement of the tables}
The ICC are presented for 104$\leq$Z$\leq$126, multipolarities 
E1 to E4 and M1 to M4, and 16 transition energies from 10 keV to
1 MeV. For any Z, the coefficients on all subshells as well as
the total ICC are tabulated.

The tables are presented as a set of independent files. For any element,
Z=nnn, there is a single file having the name {\tt nnn.TXT}
and containing the ICC
relevant to that Z as a pure ASCII text. The text spans over 74 columns.
We have chosen this 
presentation since it can be easily watched on the screen or, if necessary,
printed. Moreover, it is most of all suitable for 
operation-system-independent machine processing.

Organization of particular files is self-explaining. Nevertheless
we present here a short description.

As a heading, the atomic number is given. Then the subshell 
occupation numbers used to determine the total ICC are presented.
Next, blocks corresponding to a particular multipolarity and
six particular subshells are repeated. In Table 1, one such
 block --- for the multipolarity E1 --- is shown as an example.
\begin{table}[h]
\begin{center}
\caption{Part of the ICC for Z=104 and multipolarity E1}
\vspace*{7mm}
 \label{t:tab1}
\begin{tabular}{|rcccccc|}
\multicolumn{4}{l}{Multipolarity E1}\\
\hline
eng[keV]  & K &  L1 &  L2 & L3 & M1 &  M2    \\
\hline
  10.00 &   ***$^{a)}$  &  ***  &  ***  &  ***  & 5.727E-01 & 1.095E+00\\
  15.00 &    *** &  ***  &  ***  &  ***  & 3.545E-01 & 6.630E-01\\
  20.00 &  ***  &  *** &  *** & *** &  2.343E-01 &  4.093E-01\\
 .....&\ &\ &\ &\ &\ &\  \\
 .....&\ &\ &\ &\ &\ &\  \\
 800.00 & 6.957E-03 & 1.100E-03 & 2.376E-04 & 7.478E-05 & 2.613E-04 
   & 6.365E-05\\
1000.00 & 4.916E-03 & 7.643E-04 & 1.421E-04 & 4.138E-05 & 1.814E-04
 &  3.821E-05\\
\hline
\end{tabular}\\
$^{a)}$  three asterisks mean the energy below the threshold
   for the particular subshell
\end{center}
\end{table}
It is the first one in each file. Then the same blocks follow for
the other multipolarities. After exhausting all multipolarities,
an analogical octet of blocks follows for the next six subshells etc.
The values of total ICC are presented formally as the last
subshell ones.

\vspace{5mm}

{\it This work was supported by Grant Agency
of Czech Republic under the contract No. 202/00/1625.}

\end{document}